
%
%
\input harvmac
\def\rhob{{\rho\kern-0.465em \rho}}

\def\no{\noindent}
\def\o{\over}

\def\nl{\hfill\break}
\def\alp{\alpha}\def\gam{\gamma}\def\del{\delta}\def\lam{\lambda}

\def\ZZ{{\bf Z}}

\def\ontopss#1#2#3#4{\raise#4ex \hbox{#1}\mkern-#3mu {#2}}

\setbox\strutbox=\hbox{\vrule height12pt depth5pt width0pt}

\def\strut{\relax\ifmmode\copy\strutbox\else\unhcopy\strutbox\fi}

\nref\rbpz{A.A. Belavin, A.M. Polyakov and A.B. Zamolodchikov, J.
 Stat. Phys. 34 (1984) 763, and Nucl. Phys. B241 (1984) 333.}
\nref\rcar{J.L.~Cardy, Nucl.~Phys.~B270 (1986) 186.}
\nref\rchial{E. Verlinde, Nucl. Phys. B300 (1988) 360;\nl
 G. Moore and N. Seiberg, Commun. Math. Phys. 123 (1989) 177.}
\nref\rGKO{P.~Goddard, A. Kent and D. Olive, Commun. Math. Phys. 103
 (1986) 105.}
\nref\rkp{V.G. Kac and D.H. Peterson, Adv. in Math. 53 (1984) 125.}
\nref\rjm{M. Jimbo and T. Miwa, Adv. Stud. in Pure Math. 4 (1984) 97.}
\nref\rKaWa{V.G.~Kac and M.~Wakimoto, Adv.~in Math.~70 (1988) 156.}
\nref\rbcn{H.W. Bl{\" o}te, J.L. Cardy and M.P. Nightingale, Phys. Rev.
 Lett. 56 (1986) 742.}
\nref\raff{I. Affleck, Phys. Rev. Lett. 56 (1986) 746.}
\nref\rISZ{C. Itzykson, H. Saleur and J.-B. Zuber, Europhys. Lett. 2
 (1986) 91.}
\nref\rKob{N. Koblitz, {\it Introduction to elliptic curves and modular
 forms} (Springer, New York, 1984).}
\nref\rKac{V.G. Kac, {\it Infinite dimensional Lie algebras}
 (Cambridge University Press, 1985).}
\nref\rFF{B.L Feigin and D.B. Fuchs, Funct. Anal. Appl. 17 (1983) 241.}
\nref\rF{G. Felder, Nucl. Phys. B317 (1989) 215.}
\nref\rFS{D. Friedan and S.H. Shenker, Nucl. Phys. B281 (1987) 509.}
\nref\rRoCa{A.~Rocha-Caridi, in: {\it Vertex Operators in Mathematics
 and Physics}, ed. J.~Lepowsky {\it et al} (Springer, Berlin, 1985).}
\nref\rFQS{D. Friedan, Z. Qiu and S.H. Shenker, Phys. Rev. Lett. 52 (1984)
 1575.}
\nref\rciz{A.~Cappelli, C.~Itzykson and J.-B.~Zuber, Nucl.~Phys.~B280 (1987)
 445;\nl D.~Gepner, Nucl.~Phys.~B287 (1987) 111.}
\nref\rHuse{D.A. Huse, Phys. Rev. B30 (1984) 3908.}
\nref\rABF{G.E. Andrews, R.J. Baxter and P.J. Forrester, J. Stat. Phys.
  35 (1984) 193.}
\nref\rabs{D. Altschuler, M. Bauer and H. Saleur, J. Phys. A23 (1990) L789.}
\nref\rZF{A.B. Zamolodchikov and V.A. Fateev, Sov. Phys. JETP 62 (1985)
 215.}
\nref\rdq{J. Distler and Z. Qui, Nucl. Phys. B336 (1990) 533.}
\nref\rDJKMO{M. Jimbo, T. Miwa and M. Okado, Mod. Phys. Lett. B1 (1987) 73;\nl
 E. Date, M. Jimbo A. Kuniba, T. Miwa and M. Okado,
 Nucl. Phys. B290 (1987) 231;\nl
 M. Jimbo, T. Miwa and M. Okado, Nucl. Phys. B300 (1988) 74.}
\nref\rGcosetchar{D. Kastor, E. Martinec and Z. Qiu, Phys. Lett. 200B (1988)
 434; \nl J. Bagger, D. Nemeshansky and S. Yankielowicz, Phys. Rev. Lett.
 60 (1988) 389; \nl
 P.~Christe and F.~Ravanini, Int.~J.~Mod.~Phys.~A4  (1989) 897.}
\nref\rFL{V.A. Fateev and S.L. Lykyanov, Sov. Sci. Rev. A Phys. 15 (1990) 1.}
\nref\rJMOcmp{M. Jimbo, T. Miwa and M. Okado, Commun. Math. Phys.
 116 (1988) 507.}
\nref\rBR{V.V.~Bazhanov and N.Yu.~Reshetikhin,
 Int. J. Mod. Phys. A4 (1989) 115, and J.~Phys.~A23 (1990)
 1477.}
\nref\rKun{A. Kuniba, Nucl. Phys. B389 (1993) 209.}
\nref\rlp{J.~Lepowsky and M.~Primc, {\it Structure of the
 standard modules for the affine Lie algebra $A_1^{(1)}$},
 Contemporary Mathematics, Vol.~46 (AMS, Providence, 1985).}
\nref\rrogone{L.J. Rogers, Proc. London Math Soc. 25 (1894) 318.}
\nref\rrogtwo{L.J. Rogers, Proc. Cambridge Phil. Soc. 19 (1919) 211.}
\nref\rram{S. Ramanujan, Proc. Cambridge Phil. Soc. 19 (1919) 214.}
\nref\rgor{B. Gordon, Amer. J. Math. 83 (1961) 393.}
\nref\rand{G.E. Andrews, Proc. Nat. Acad. Sci. USA 71 (1974) 4082.}
\nref\ral{G. Albertini, J. Phys. A25 (1992) 162.}
\nref\rADMone{G. Albertini, S. Dasmahapatra and B.M. McCoy, Int. J.
 Mod. Phys. A7, Suppl. 1A (1992) 1.}
\nref\rADMtwo{G. Albertini, S. Dasmahapatra and B.M. McCoy, Phys. Lett.
 170A (1992) 397.}
\nref\rKedMc{R.~Kedem and B.M.~McCoy, J. Stat. Phys. (in press).}
\nref\rDKMM{S. Dasmahapatra, R. Kedem, B.M. McCoy and E. Melzer
 (in preparation).}
\nref\rAMP{G. Albertini, B.M. McCoy and J.H.H. Perk, Phys. Lett.
 135A (1989) 159.}
\nref\rBBP{R.J. Baxter, V.V. Bazhanov and J.H.H. Perk, Int. J.
 Mod. Phys. B4 (1990), 803.}
\nref\rPearce{P.A. Pearce, Int. J. Mod. Phys. A7, Suppl. 1B (1992) 791.}
\nref\raltwo{G. Albertini, S. Dasmahapatra and B.M. McCoy (these
 proceedings).}
\nref\rDots{Vl.S. Dotsenko, Nucl. Phys. 235 (1984) 54.}
\nref\rgq{D. Gepner and Z. Qiu, Nucl. Phys. B285 [FS19] (1987) 423.}
\nref\rPeartwo{P.A. Pearce and X.K. Zhao (these proceedings).}
\nref\rstan{R.P. Stanley, {\it Ordered structures and partitions}, Mem.
 Amer. Math. Soc. 119 (1972).}
\nref\rKKMMone{ R. Kedem, T.R. Klassen, B.M. McCoy and E. Melzer,
 Phys. Lett. B (in press).}
\nref\rKKMMtwo{R.Kedem, T.R. Klassen, B.M. McCoy, and E. Melzer,
 Phys. Lett. B (submitted).}
\nref\rGinsp{P. Ginsparg, Nucl. Phys. B295 (1988) 153.}
\nref\rTer{M.~Terhoeven, Bonn preprint BONN-HE-92-36, hep-th/9111120.}
\nref\rKNS{A. Kuniba, T. Nakanishi, J. Suzuki, Harvard preprint HUPT-92/A069.}
\nref\rgeptwo{D. Gepner, Nucl. Phys. B290 (1987) 10.}
\nref\rFNO{B.L. Feigen, T. Nakanishi and H. Ooguri, Int. J. Mod.
 Phys. A7, Suppl. 1A (1992) 217.}
\nref\rNRT{W. Nahm, A. Recknagel and M. Terhoeven, Bonn preprint,
 hep-th/9211034.}
\nref\rgepth{D. Gepner, Nucl. Phys. B296 (1988) 757.}
\nref\rRS{B. Richmond and G. Szekeres, J. Austral. Soc. (Series A) 31
 (1981) 362.}
\nref\rLewin{L. Lewin, {\it Dilogarithms and associated functions}
 (MacDonald, London, 1958).}
\nref\rKR{A.N.~Kirillov and N.Yu.~Reshetikhin, J.~Phys.~A20 (1987) 1587.}
\nref\rKir{A.N.~Kirillov, Zap.~Nauch.~Semin.~LOMI 164 (1987) 121
 (J.~Sov.~Math.~47 (1989) 2450), and Cambridge preprint, hep-th/9212150.}
\nref\rKlaMel{T.R.~Klassen and E.~Melzer, Nucl.~Phys.~B338 (1990) 485.}
\nref\rFatZam{V.A. Fateev and Al.B. Zamolodchikov, Phys. Lett. 271B (1991) 91.}
\nref\rKlumPear{A.~Kl\"umper and P.A.~Pearce, J.~Stat.~Phys.~64
 (1991) 13; Physica A 183 (1992) 304.}
\nref\rKunNak{A. Kuniba and T. Nakanishi, Mod. Phys. Lett. A7 (1992) 3487.}
\nref\rFrenk{E. Frenkel and A. Szenes, Harvard preprint, hep-th/9212094.}
\nref\rTBA{V.M. Filyov, A.M. Tsvelik and P.B. Wiegman, Phys. Lett. 81A
 (1981) 175;\nl
 H.M. Babujian, Nucl. Phys. B215 (1983) 317;\nl
 R. Kedem, Stony Brook preprint ITP-SB-92-60, J.~Stat.~Phys. (in press);\nl
 L. Mezincescu, R.I. Nepomechie, P.K. Townsend and A.M.Tsvelik, Miami
 preprint UMTG-169 (1992).}

\Title{\vbox{\baselineskip12pt\hbox{ITP-SB-93-12, RU-93-07}
\hbox{hep-th/9303013}}}
{\vbox{\centerline{Quasi-Particles, Conformal Field Theory, and $q$-Series}}}

\vskip 6mm
\centerline{S.~Dasmahapatra,\foot{ICTP, Strada Cortiera 11, I34100
 Trieste, Italy}~
 R.~Kedem,\foot{Institute for Theoretical Physics,
 SUNY, Stony Brook,  NY 11794-3840}~
 T.R.~Klassen,\foot{Department of Physics and Astronomy, Rutgers
 University, Piscataway, NJ 08854-0849} ~B.M.~McCoy,$^2$~ and ~E.~Melzer$^2$}

\vskip 17mm

\centerline{\bf Abstract}
\vskip 3mm

We review recent results concerning the representation of
conformal field theory  characters
in terms of fermionic
quasi-particle excitations, and
describe in detail their construction in the case of the
integrable three-state  Potts chain.
These fermionic representations are $q$-series which are generalizations
of the sums occurring in the Rogers-Ramanujan identities.

\vskip 23mm
To appear in the proceedings of ``Yang-Baxter Equations in Paris'',
July 1992, J.-M.~Maillard (ed.).

\Date{\hfill 2/93}
\vfill\eject

\newsec{Introduction}

The fundamental problem of all condensed matter physics is the
explanation of macroscopic phenomena in many-body systems in terms of
a microscopic quantum mechanical description of the system. For all
practical applications there is no dispute that the microscopic
description of the world is as a collection of electrons
and nuclei which interact by electromagnetic forces (which may usually
be well thought of as non-relativistic Coloumb interactions and
possibly a spin-orbit coupling). The problem is to extract macroscopic
collective properties from this microscopic interaction.

The importance and difficulty of this problem is revealed in the
question of the origins of organic chemistry. All organic molecules of
biological significance, such as DNA, are optically active and rotate
light in a preferred direction. This rotation clearly violates parity.
Nevertheless, the underlying microscopic interaction is parity
invariant. This vividly illustrates the fact that the physics of the
collective excitations may be qualitatively different from that of the
underlying microscopic system.

It is thus no surprise that the study of collective excitations in
macroscopic systems is far from understood. It is also not surprising
that approximate methods have only limited utility in building
insight into these phenomena. Thus it is that ever since the invention
of quantum mechanics there has been constant attention to the
problem of finding and studying simplified microscopic model systems
for which exact, nontrivial computations can be done which give insight
into the relation of the collective to the microscopic.

In this paper
we will discuss two such approaches which have proven exceedingly
fruitful: integrable models of statistical mechanics
and conformal field theory. We
will discuss these in relation to what is one of the most simple of
macroscopic properties: the low-temperature behavior of the specific
heat. It is one of the loveliest discoveries of the past decade that
this most simple of collective properties has profound connections to
the theory of representations of affine Lie algebras and the
mathematical study of $q$-series and generalized Rogers-Ramanujan
identities.

\bigskip
\newsec{Specific Heat and Quasi-Particles}

Perhaps the most fundamental quantity used in the study of macroscopic
systems is the partition function defined as
\eqn\part{Z=\Tr\ e^{-{H/ {k_BT}}}}
where $H$ is the hamiltonian, the trace is
over all states of the system, $k_B$ is Boltzmann's constant
and $T$ is the temperature. More explicitly this may be written as
\eqn\Zsum{Z=e^{-{E_{GS}/ k_BT}}\sum_ne^{-(E_n-E_{GS})/k_BT}}
where the sum is over all the eigenvalues $E_n$ of $H$ and we have
explicitly factored out the contribution of the ground state
energy $E_{GS}$.

For a macroscopic system we are usually more interested in the free
energy per site $f$ in the thermodynamic limit, defined as
\eqn\free{
f=-k_BT\lim_{M \rightarrow \infty}{1\over M}\ln Z~,}
where $M$ is the size of the system, and for concreteness we will
think of $H$ as the hamiltonian of a spin system of  a linear chain of
$M$ sites. The thermodynamic limit is defined as
\eqn\TM{{\rm fixed}\quad T>0\quad {\rm \quad and\quad \quad}
 M \rightarrow \infty~,}
and the specific heat is given as
\eqn\sheat{C~=~-T~{\partial^2 f\over{\partial T^2}}~~.}
The low-temperature behavior of the specific heat is now obtained by
taking $T \rightarrow 0.$

To evaluate the sum~\Zsum\ and thus to study the specific heat~\sheat\
we need to study the energy levels of $H$ (which are obtained from
Schr\"odinger's equation) in the $M \rightarrow \infty$ limit. We will
further make the assumption that $H$ is translationally invariant with
periodic boundary conditions so that the momentum $P$ is a good
quantum number. It is then almost universally found that if $E-E_{GS}$
is finite and non-zero as $M \rightarrow \infty$ then the energy
levels may be expressed in terms of single-particle levels
$e_{\alpha}(P_i^\alpha)$, with $\alpha$ labelling the type of
excitation, which depend on a momentum $P_i^\alpha$ and a set of
combination rules as
\eqn\Eqp{E-E_{GS}~=~\sum_{\alpha,{\rm rules}}~~\sum_{i=1}^{m_\alpha}
   e_{\alpha}(P_i^\alpha)~,}
and that the total momentum is given as
\eqn\mom{P~\equiv~\sum_{\alpha,{\rm rules}}~~\sum_{i=1}^{m_\alpha}
   P_i^{\alpha}~({\rm mod}~2\pi).}
Energy levels of a many-body system of this form are said to be a
quasi-particle spectrum. When one of the rules of composition is the
fermi exclusion rule
\eqn\fermi{P_i^{\alpha} \neq P_j^{\alpha} \quad \quad {\rm if}\quad
 \quad  i \neq j~,}
the spectrum is said to be fermionic.

If $e_{\alpha}(P)$ is positive
for all $P$ the system is said to have a mass gap, and the specific heat
vanishes exponentially as $T\rightarrow 0$.
 However, in many spin
chains one or more $e_{\alpha}(P)$ vanish as $P \rightarrow 0$ as
\eqn\elin{e(P)~\sim ~v|P|~,}
 where $v$ is positive. These systems are said to be massless and $v$ is
called the speed of sound. If this massless single-particle energy is
used in~\Eqp\ and \Zsum\ and the momenta $P_i$ are taken to have a
uniform distribution it is a familiar result that (with a single
species of excitation) the specific heat vanishes linearly when $T
\rightarrow 0$ as
\eqn\lt{C~\sim ~{\pi k_B{\tilde c}\over 3v}~T~,}
where ${\tilde c}$ is a constant which is equal
to ${1\over 2}$ in this case.

This argument, however, is not complete
as is apparent from the
observation that any energy level with $\lim_{M\to \infty}e(P)>0$
will contribute only a term exponentially small in $T$ to the
specific heat. Thus the order one excitations which are of the
form \Eqp\ do not contribute to the linear behavior~\lt. Instead, it
is the levels with the property that $\lim_{M\rightarrow
\infty}e(P)=0$ which contribute to the leading behavior.

\bigskip
\newsec{Conformal Field Theory}

In contrast to the condensed matter description of quasi-particles of
the previous section, the study of the $1\over M$ excitation energies is
much more recent and, in particular, the most remarkable progress has
been made only in the last decade starting with the seminal work of
Belavin, Polyakov and Zamolodchikov~\rbpz~on conformal field theory.

In the statistical mechanics context, the work of~\rbpz~applies directly
to the continuum limit of two-dimensional lattice models which
are assumed to exhibit conformal invariance at criticality.
A fundamental object~\rcar~in that framework is the
finite-size (classical)
partition function $\hat{Z}_{2d}$ of the critical system.
Namely, consider the partition function at $T_c$ of the
(possibly anisotropic) system
on an $M$ by $M'$ periodic lattice
\eqn\Ztwod{ Z_{2d}(M,M')= \sum_{{\rm states}} e^{-{\cal E}/k_B T_c}
 = \sum_j (\Lambda_j(M))^{M'}~,}
which we expressed in terms of the eigenvalues $\Lambda_j(M)$
of the transfer matrix ${\cal T}_M$
in one of the directions. Defining the bulk free energy
$f_{2d} \equiv -k_B T_c \lim_{M,M'\to\infty}{1\o MM'} \ln Z_{2d} =
-k_B T_c \lim_{M\to\infty} {1\o M}\ln \Lambda_{{\rm max}}(M)$,
the finite-size partition function is defined by scaling out the bulk free
energy via
\eqn\Zhatwod{ \hat{Z}_{2d}
  = \lim_{M,M'\to\infty} e^{MM'f_{2d}/k_B T_c} Z_{2d}
 = \lim_{M,M'\to\infty}
 \sum_j \left({\Lambda_j(M)\o \Lambda_{\rm max}(M)}\right)^{M'}~.}
the limit being taken with $M'/M$ held fixed. $\hat{Z}_{2d}$ is a finite
function of $q_{2d} = e^{\alpha M'/M}$, where $\alpha$
(possibly complex) depends on the anisotropy.

The analogous object in the context of the gapless spin chain  which is of
interest to us here, is the (quantum) partition function \Zsum\ in the limit
\eqn\clim{M\rightarrow \infty,\quad T \rightarrow 0
 \quad \quad{\rm with\quad }\quad MT
\quad {\rm fixed,}}
which focuses directly on the order ${1 \over M}$ energy
levels of the hamiltonian. More precisely, introducing
$e_0 \equiv \lim_{M\to\infty}{1\o M}E_{GS}$, define
\eqn\Zhat{ \hat{Z}=\lim e^{Me_0/k_B T} Z}
in the limit \clim, so that $\hat{Z}$ is a finite function of
\eqn\qdef{q=\exp\left({\textstyle -{2 \pi v\over Mk_B T}} \right)~. }
For a hamiltonian obtained from a family of commuting transfer matrices
${\cal T}_M(u)$ of an integrable critical lattice model via
$H={d\o du}\ln {\cal T}_M(u)\big|_{u=u_0}$, where $u_0$ is a special value
of the spectral parameter where ${\cal T}_M$ becomes the identity,
$\hat{Z}$ coincides as a function with the corresponding $\hat{Z}_{2d}$.

The limit~\clim\ is not the same as the limit~\TM\ which defines the specific
heat. However, if no additional length scale appears in the system, it
is expected that the behavior of the specific heat computed using the
prescription~\clim\ will
agree when $q \rightarrow 1$ with the $T \rightarrow 0$ behavior
computed using the prescription~\TM.
We are therefore led to discuss the $q\to 1$ behavior of $\hat{Z}(q)$.

\medskip
An important feature of conformal field theory
is~\rcar~that $\hat{Z}$ can be expressed in the factorized form
\eqn\zfac{ \hat{Z}(q)~=~\sum_{k,l} N_{kl}~\chi_{k}(q)~\chi_{l}(\bar{q})~,}
where the $\chi_k(q)$ are characters of a chiral algebra~\rchial , with
the $N_{kl}$ non-negative integers.
(In so-called coset models of
conformal field theory~\rGKO , the characters are known to be branching
functions~\rkp\rjm\rKaWa~of some affine Lie algebras.)
In the two-dimensional context $\bar{q}$ in \zfac\ is the complex conjugate
of $q$, while in the one-dimensional one $q$ and $\bar{q}$ are real and
equal and are associated with contributions from right- and left-movers,
respectively.
We will restrict attention to rational conformal field theories,
where the sum in \zfac\ is finite.
The characters take the form
\eqn\chihat{ \chi_k(q) = q^{\Delta_k-{c\o 24}} \hat{\chi}_k(q)~~,
 ~~~~~~~~~\hat{\chi}_k(q) = 1 + \sum_{n=1}^\infty a_n q^n~~,}
with the $a_n$ non-negative integers. Here $c$
and the $\Delta_k$ are the central charge and conformal dimensions,
respectively, of the conformal field theory.

The partition function $\hat{Z}_{2d}$ of the two-dimensional
system must clearly have the property
\eqn\Sinv{ \hat{Z}_{2d}(q)=\hat{Z}_{2d}(\tilde{q})~,}
where
\eqn\qtilde{ \tilde{q}=e^{-2\pi i/\tau}~~~~~~~{\rm when}~~~~~~
  q=e^{2\pi i\tau}~,}
simply by symmetry in $M$ and $M'$ combined with an appropriate change in
the anisotropy, when present. If \Sinv\ holds for
$\hat{Z}$ as well, then one concludes from \zfac\ and \chihat\ that
\eqn\Zasymp{ \hat{Z}(q)~ \sim ~\tilde{q}^{-(c-12d_{\rm min})/12}~~~~~~~
 {\rm as} ~~~q\to 1^- ~,}
where $d_{\rm min}$ is the minimal $\Delta_k+\Delta_l$ such that $N_{kl}>0$.
This shows that the $q\to 0$ behavior of $\hat{Z}(q)$, determining
the finite-size corrections to the ground state energy~\rbcn \raff\rISZ~
\eqn\fsc{ -k_B \lim_{T\to 0} T \ln Z ~=~ E_{GS}-Me_0~ =~
 -{\pi (c-12d_{\rm min})v \o 6M} + o(M^{-1})~,}
is related to the $q\to 1$ behavior which is relevant for the specific
heat. Namely,
from \Zasymp\    and \qdef\
we conclude that
$\tilde{c}$ in \lt\ is given by
\eqn\ctilde{ \tilde{c} = c-12d_{\rm min}~, }
where the rhs is called the effective central charge.

\medskip
One of the objectives in this work is to point out an alternative
method to compute the low-temperature specific heat, which is based
on an analysis of the order one energy levels of the hamiltonian
and bypasses the use of \Sinv\ which is a property not a priori obvious
from the
viewpoint of a generic one-dimensional chain. We will demonstrate (in two
particular models) how the full partition function $\hat{Z}$ --- or
at least the ``normalized characters'' $\hat{\chi}_k(q)$ --- can be
obtained from the quasi-particle description of the spectrum
discussed in sect.~2 (where the specifics of a model are encoded
in the ``rules'' in \Eqp).
The specific heat is then deduced
using \zfac\
from the $q\to 1$ behavior
of the $\hat{\chi}_k(q)$,
which can be determined by a steepest descent calculation.
The leading behavior, which is the same for all characters in a given
model, is
\eqn\chiasy{ \hat{\chi}_k(q) ~\sim ~\tilde{q}^{-\tilde{c}/24}~~~~~~~{\rm as}
  ~~~q\to 1^- ~,}
where $\tilde{c}$     agrees with \ctilde.

\medskip
Let us emphasize that in this
computation of the low-temperature specific heat no
use
of modular covariance~\rKob~of the characters is made.
The approach pioneered in~\rbpz~relies
on the existence
of conformal symmetry in the system, which severely constrains
the order ${1\o M}$ spectrum in terms of representations of some
infinite-dimensional chiral algebra~\rchial.
The characters of these chiral algebra representations
are computed either abstractly~\rKac~or by
the Feigin-Fuchs-Felder
construction~\rFF \rF. Using these methods, the explicit
expressions obtained for the characters usually involve
modular forms, and therefore the
characters ${\chi}_k(q)$ of a given model
(regarded as functions of a complex variable $q$)
can be seen to form~\rcar \rkp \rFS~a representation of
the modular group,
generated by ~$S$:~$q\to \tilde{q}$~ and
{}~$T$:~$q\to e^{2\pi i}q$.
In particular, they satisfy a linear transformation law
\eqn\modco{ \chi_k(\tilde{q}) ~=~ \sum_l S_{kl} ~\chi_l(q) ~,}
from which \chiasy\ can be obtained.

However, the detailed connection between the above-mentioned
expressions for the
characters
in terms of modular forms
to a hamiltonian spectrum is
rather obscure.
In the approach of this paper, alternative expressions --- which we call
(fermionic) quasi-particle
representations --- for the
characters
are obtained
from the spectrum, and thus a direct understanding of the
conformal field theory partition function $\hat{Z}$ in terms of
the underlying spin chain is gained.

\bigskip

We will now provide some more details in a few examples.
In the past 10 years there has been an immense effort to discover and
classify conformal field theories, compute the
corresponding characters and partition functions, and identify
the underlying statistical mechanics models. The earliest example
is the series of
minimal models ${\cal M}(p,p')$~\rbpz,
specified by pairs of coprime positive integers
$p$ and $p'$, where the central charge is
\eqn\ccharge{c=1-{6(p-p')^2\over p p'}}
and the conformal dimensions are
\eqn\del{\Delta_{~r,s}^{(p,p')}={(rp'-sp)^2-(p-p')^2\over 4p p'}
  ~~~~~~~(r=1,\ldots,p-1;~~s=1,\ldots,p'-1).}
The corresponding characters are~\rFF \rF \rRoCa
\eqn\roc{q^{c/24}\chi_{~r,s}^{(p,p')}={q^{\Delta_{~r,s}^{(p,p')}}\over
  (q)_{\infty}}
  \sum_{k=-\infty}^{\infty}(q^{k(kpp'+rp-sp')}-q^{(kp'+s)(kp+r)})}
where
\eqn\q{(q)_{n}=\prod_{k=1}^n (1-q^k)~.}
The unitary~\rFQS~minimal conformal field theories ${\cal M}(p,p+1)$
(with the $A$-series partition
function~\rciz) were identified~\rHuse~as describing the
continuum limit of the
RSOS models of Andrews, Baxter and Forrester~\rABF~at the critical point
between regimes III and IV.

A second widely studied class of theories comprises the coset models~\rGKO
\eqn\coset{{(G_r^{(1)})_k\times (G_r^{(1)})_l\over (G_r^{(1)})_{k+l}}~~,}
where $(G_r^{(1)})_k$ is the affine Lie algebra at level $k$~\rKac~based
on the simply-laced Lie algebra $G_r$ of rank $r$.
(The unitary minimal models ${\cal M}(p,p+1)$ are obtained~\rGKO~from
\coset\ by specializing to $G_r=A_1$, $k=p-2$, and $l=1$.)
For the case $k=l=1$ and $G_r=A_{N-1}$ the model~\coset\
is identical by level-rank duality~\rabs~to the coset model
${(A_1^{(1)})_N\over U(1)}$,
known as $\ZZ_N$-parafermionic conformal field theory~\rZF .
The central charge is
\eqn\cchargepf{c={2(N-1)\over N+2}~~,}
and the characters are branching functions given by Hecke indefinite
forms of~\rkp\rjm~ (or an equivalent form~\rdq )
\eqn\blm{\eqalign{
 q^{c/24}b_m^l & = \cr  {q^{h_m^l}\over (q)_{\infty}^2}&
 \Bigg[\Bigg(\sum_{s\geq0}\sum_{n\geq0}-\sum_{s<0}\sum_{n<0}\Bigg)
 (-1)^s q^{s(s+1)/2+(l+1)n+(l+m)s/2+(N+2)(n+s)n}\cr
 +& \Bigg( \sum_{s>0}\sum_{n\geq0}-\sum_{s\leq0}\sum_{n<0}\Bigg)
 (-1)^s q^{s(s+1)/2+(l+1)n+(l-m)s/2+(N+2)(n+s)n}\Bigg]~,\cr}}
where the dimensions $h^l_m$ are
\eqn\hlm{h_m^{l}={l(l+1)\over 4(N+2)}-{m^2\over 4N}~~.}
Here $l=0,1,\ldots,N-1$, ~$l-m$ is even, and the formulas are valid for
$|m| \leq l$ while for $|m|>l$ one uses the symmetries
\eqn\sym{b_m^l=b_{-m}^l=b_{m+2N}^l=b_{N-m}^{N-l}~.}
For the more general cosets of~\coset~the branching functions can
be found in~\rDJKMO \rGcosetchar \rFL .
The statistical mechanical models underlying the theories \coset\ are
discussed in~\rDJKMO\rJMOcmp\ \rBR\rKun .

The above expressions for the characters, from which their modular
properties can be derived,
all have the feature that there are
several powers of $(q)_{\infty}$ in the denominator,
corresponding to
the fact that the Feigin-Fuchs-Felder construction
from which
they can be obtained  is based on bosonic Fock spaces (which are
then truncated in a particular way, encoded by the ``numerator'').
We will call such representations bosonic.

But there are other
forms in which the characters
may be expressed. Most notable is the equivalent form of the branching
functions~\blm~obtained by Lepowsky and Primc~\rlp
\eqn\lpsum{
 q^{c/24}b_{2Q-l}^l~=~q^{{l(N-l)\over 2N(N+2)}}
 \sum_{\scriptstyle m_1,\ldots ,m_{N-1}=0\atop \scriptstyle {\rm restrictions}}
 ^\infty {q^{{\bf m}C_{N-1}^{-1}{\bf m}^t-{\bf A}_l\cdot{\bf m}}\over
 (q)_{m_1} \ldots (q)_{m_{N-1}}},}
where ${\bf m}=(m_1,\ldots,m_{N-1})$ is subject to the restriction
\eqn\rest{\sum_{\alpha=1}^{N-1}\alpha m_\alpha\equiv Q~({\rm mod}~N),}
$C_{N-1}$ is the Cartan matrix of the Lie algebra $A_{N-1}$
in the basis where we explicitly have
\eqn\qfAn{  {\bf m} C_{N-1}^{-1} {\bf m}^t ~=~
  {1\over N} \left( \sum_{\alpha=1}^{N-1} \alpha(N-\alpha)m_\alpha^2
 + 2\sum_{1\leq \alpha < \beta \leq N-1}
     \alpha(N-\beta) m_\alpha m_\beta \right)~,}
and
\eqn\lin{{\bf A}_l\cdot{\bf m}= -({\bf m}C_{N-1}^{-1})_l =
 -\left( {N-l\over N}\sum_{\alpha=1}^{l} \alpha m_\alpha
 +{l\over N}\sum_{\alpha=l+1}^{N-1}(N-\alpha)m_\alpha\right)~.}
This representation is of the form of a $q$-series which generalizes the
sum-side of the Rogers-Ramanujan identities~\rrogone \rrogtwo \rram~to
multiple sums,
such as appear in the Andrews-Gordon identities~\rgor \rand. For reasons
that will become clear
in the next sections we refer to such a
representation as fermionic.

\bigskip
\newsec{Three state Potts chain}

The general discussion of specific heat and quasi-particles of sect.~2
and the sketch of conformal field theory of the previous section do
not rely on any microscopic hamiltonian. There are, however, a large
number of integrable spin chains and corresponding two-dimensional
classical statistical mechanics systems which are closely related to
conformal field theories. These spin chains have eigenvalue spectra
which can be studied by means of functional and Bethe's
equations. It is thus natural to attempt to compute the
conformal field theory characters from the spin chain.

This program has recently been carried out~\ral-\rDKMM~for the 3-state Potts
chain. We will here summarize the results of this study to illustrate
the relations which  both the Rocha-Caridi~\roc~and the
Lepowsky-Primc~\lpsum\ character   formulae have to the spin chain and
to the order one excitations~\Eqp\ of condensed matter physics. This
investigation will lead to a physical interpretation of~\lpsum\ and a
new representation for~\roc.

\subsec{The hamiltonian and Bethe's equations.}

The 3-state Potts chain is specified by the hamiltonian
\eqn\ham{ H~=~{\pm{2}\over{\sqrt3}}~\sum_{j=1}^{M}\left(
 X_{j}+X_{j}^{\dagger} +
 Z_{j}Z_{j+1}^{\dagger}+Z_{j}^{\dagger}Z_{j+1}\right)~,}
where
\eqn\xzj{ X_{j}=I\otimes I\otimes \cdots \otimes {\underbrace X_{j^{th}}}
 \otimes \cdots \otimes I~,~~~~~~~~
 Z_{j}=I\otimes I \otimes \cdots \otimes {\underbrace Z_{j^{th}}}
 \otimes\cdots \otimes I~.}
Here $ I$ is the $3 \times 3$ identity matrix,
\eqn\defxz{ X=\pmatrix{0 & 0 & 1\cr 1 & 0 & 0\cr 0 & 1 & 0\cr}~~,
 ~~~~~Z=\pmatrix{1 & 0 & 0\cr 0 & \omega & 0\cr 0 & 0 & \omega^2\cr}~~,
 ~~~~~\omega=e^{2\pi i/3}~,}
and we impose periodic boundary conditions $Z_{M+1}\equiv Z_{1}$.
If the $-~(+)$ sign is chosen in \ham, the spin chain is called ferromagnetic
(anti-ferromagnetic).

This spin chain is invariant under translations and under $\ZZ_3$ spin
rotations. Thus the eigenvalues may by classified in terms of $P$, the
total momentum of the state, and $Q$, where $e^{2 \pi i Q/3}$ is the
eigenvalue of the spin rotation operator. Here $P=2 \pi n/M$ where $n$
is an integer $0\leq n \leq{M-1}$, and $Q=0,\pm 1$. Furthermore, because
$H$ is invariant under complex conjugation there is a conserved $C$
parity of $\pm1$ in the sector $Q=0$, and the sectors $Q=\pm1$ are
degenerate.

This spin chain is integrable because of its connection with the
two-dimensional 3-state Potts model at the critical point, which is
integrable. The eigenvalues of the transfer matrix satisfy functional
equations~\ral \rAMP - \rPearce~which are solved in terms of
Bethe equations~\ral
\eqn\bethe{
(-1)^{M+1}\left[{\sinh(\lambda_j-iS\gamma)\over{\sinh(\lambda_j+iS\gamma)}}
\right]^{2M}=~\prod_{k=1}^{L}~{\sinh(\lambda_j-\lambda_k-i\gamma)\over
{\sinh(\lambda_j-\lambda_k+i\gamma) }}}
with
\eqn\gam{ \gamma={\pi\o 3}~,\quad ~~~~S={1\o 4}~,~~~~~~~
L=2(M-|Q|)\quad {\rm for}\quad Q=0,\pm1~.}
In terms of these ${\lambda_k}$, the eigenvalues of the transfer
matrix of the statistical model are
\eqn\lam{
\Lambda(\lambda)=\left[{\sinh({\pi i\over6})\sinh({\pi i\over3})
\over{\sinh({\pi i\over4}-\lambda)\sinh({\pi i\over4}+\lambda)}}
\right]^{M}\prod_{k=1}^{L}~{\sinh(\lambda-\lambda_k)\over{\sinh({\pi
i\over12}+\lambda_k)}}~~,}
the eigenvalues of the hamiltonian~\ham~ are
\eqn\energy{
E=\sum_{k=1}^{L}\cot(i\lambda_k+{\pi\over{12}})-{2M\over{\sqrt3}}~~,}
and the corresponding momentum is
\eqn\momenta{
e^{iP}=\Lambda(-i\pi/12)=\prod_{k=1}^{L}~{\sinh(\lambda_k+
{\pi i\over12})\over{\sinh(\lambda_k-{\pi i\over12})}}~~.}

\subsec{Order one excitations.}
These equations have been recently solved to obtain the order one
excitation energies~\rADMtwo. The computations are discussed in
detail in the article in these proceedings~\raltwo. The
results are as follows (we describe them in detail only the sector $Q=0$):

\vskip11pt

\no
{$\bullet$~
Ferromagnetic case:}

The order one excitation energies and momenta are
\eqn\ferroe{E-E_{GS}~=~\sum_{j=1}^{m_+}e(P_j^+)~~,
 ~~~~~~~~P~\equiv~\sum_{j=1}^{m_+}P_j^+ ~({\rm mod}~2\pi),}
where
\eqn\mplus{m_+=2m_{ns}+3m_{-}+4m_{-2s}}
with $m_{ns},m_{-}, m_{-2s}$ arbitrary non-negative integers, and
\eqn\eferro{e(P_j^{+})=6\sin({P_j^+\over 2})\quad ~~~0\leq P_j^+\leq 2\pi~,
 ~~~~~P_j^+\neq P_k^+\quad{\rm for}\quad j\neq k~.}
Each state has a degeneracy~\rADMone, which for $Q=0$ is
\eqn\deg{{m_{-}+m_{-2s}\choose m_{-}}{2m_{-}+2m_{-2s}+m_{ns}\choose
  m_{ns}}~.}
The speed of sound $v$ is found to be 3, since
\eqn\ferropzero{e(P^+)~\sim ~3 |P^+| ~~~~~~~~~{\rm for}~~~~ P^+~\sim~ 0.}

\vfill \eject
\no
{$\bullet$~ Anti-ferromagnetic case:}

We restrict our attention to $M$ even.
The order one excitation energies   and momenta in the sector $Q=0$  are
\eqn\antiferroe{E-E_{GS}~=
\sum_{\alpha=2s,-2s,ns}~~ \sum_{j=1}^{m_{\alpha}}e_{\alpha}(P^\alpha_j)~,
{}~~~~P-P_{GS}~=\sum_{\alpha=2s,-2s,ns}~~\sum_{j=1}^{m_{\alpha}}P_j^{\alpha}~,}
where $P$ is defined modulo $2\pi$, and
\eqn\pzero{P_{GS}\equiv {M\over2}\pi~({\rm mod}~2\pi)~,}
\eqn\mres{m_{2s}+m_{-2s}\quad {\rm is \quad even}.}
The single-particle momenta are subject to the fermi exclusion
rule \fermi, and the single-particle energies are
\eqn\antifenone{ \eqalign{
  e_{2s}(P)~ &=~3\{\sqrt2 \cos ({{|P|}\over 2}-{{3 \pi}\over 4})+1\}
  \quad ~~~~~~~~~~0\leq P\leq 3\pi \cr
   e_{-2s}(P)~ &=~3\{\sqrt2 \cos({{|P|}\over 2}-{{\pi}\over 4})-1\}
   \quad ~~~~~~~~~~~0\leq P \leq \pi \cr
   e_{ns}(P)~ &=~3 \sin({{|P|}\over 2})
   \quad ~~~~~~~~~~~~~~~~~~~~~~~~~~~~~~0 \leq P \leq 2\pi~.\cr} }
The speed of sound is ${3\o 2}$ for all three excitations:
\eqn\antipzero{e_{\alpha}(P)~\sim~{3\over 2}|P|~~
  ~~~~~~~~{\rm for}~~~~~P~\sim~ 0~.}

\subsec{Conformal field theory predictions.}

We turn now to the
conformal field theory predictions for the partition
functions of both the ferromagnetic and the anti-ferromagnetic cases.

\vskip11pt
\no
{$\bullet$~ Ferromagnetic case:}

The conformal field theory in this case
was identified by Dotsenko~\rDots~to be the minimal model
${\cal M}(5,6)$ of central charge $c={4\o 5}$~ (cf.~\ccharge),
and the   partition function was argued by Cardy~\rcar~to
be the modular-invariant non-diagonal combination of characters
\eqn\ztsp{\eqalign{\hat{Z}_F
{}~ = ~&[\chi_0(q)+\chi_3(q)][\chi_0(\bar q)+\chi_3(\bar q)]
 +[\chi_{2/5}(q)+\chi_{7/5}(q)][\chi_{2/5}(q)+\chi_{7/5}(\bar q)]\cr
 &+2\chi_{1/15}(q)\chi_{1/15}(\bar q)+2\chi_{2/3}(q)\chi_{2/3}(\bar q)~.\cr}}
Here we use the notation ~$\chi_\Delta = \chi^{(5,6)}_{\Delta_{r,s}}$ ~for
the characters whose first few terms
are obtained from~\roc\ as
\eqn\ctspf{\eqalign{
q^{c\o 24}\chi_0 &= 1 +  q^2 + q^3 + 2q^4 + 2q^5 + 4q^6 + 4q^7 + 7q^8
+ 8q^9 +12q^{10} \ldots \cr
q^{c\o 24}\chi_{2/5} &=  q^{2\o 5}(1 + q + q^2 + 2q^3 + 3q^4 + 4q^5 + 6q^6 +
 8q^7 +     11q^8 + 15q^9  \ldots)
\cr
q^{c\o 24}\chi_{7/5} &=q^{7\o 5}(1 + q+ 2q^2 + 2q^3 + 4q^4 + 5q^5 +
8q^6 +10 q^7 +
     15q^8 + 19q^9   \ldots)
\cr
q^{c\o 24}\chi_3     &=q^3(1 + q + 2q^2 + 3q^3 + 4q^4 + 5q^5 + 8q^6 +
10q^7 + 14q^8 + 18q^9   \ldots)
\cr
q^{c\o 24}\chi_{1/15}&= q^{1\o 15}(1 + q + 2q^2 + 3q^3 + 5q^4 + 7q^5 +
10q^6 + 14q^7 +
     20q^8 + 26q^9   \ldots)\cr
q^{c\o 24}\chi_{2/3} &= q^{2\o 3}(1 + q + 2q^2 + 2q^3 + 4q^4 + 5q^5 +
8q^6 + 10q^7 +15q^8 + 19q^9 \ldots).\cr}}

\vskip11pt
\no
{$\bullet$~ Anti-ferromagnetic case:}

In this case the conformal field theory was identified by
Pearce~\rPeartwo~to be that of $\ZZ_4$ parafermions, of central
charge $c$=1 (cf.~\cchargepf~with $N$=4),
with the non-diagonal partition function~\rgq
\eqn\ztspa{
\hat{Z}_{AF}=[b_0^0(q)+b_4^0(q)][b_0^0(\bar q)+b_4^0(\bar q)]+
 4b_2^0(q)b_2^0(\bar q)+ 2b_0^2(q)b_0^2(\bar q) +2b_2^2(q)b_2^2(\bar q)~,}
in terms of the branching functions $b_m^l$, which are
obtained from~\blm~or~\lpsum\  with $N$=4 as
\eqn\ctspa{\eqalign{
q^{1\o 24}b_0^0 & = (
 1+q^2+2q^3+4q^4+5q^5+9q^6+12q^7+19q^8+25q^9+37q^{10}\ldots)\cr
q^{1\o 24}b_2^0 & =
q^{3\o 4}(1+q+2q^2+3q^3+5q^4+7q^5+12q^6+16q^7+24q^8+33q^9\ldots ) \cr
 q^{1\o 24}b_4^0&
=q(1+q+3q^2+3q^3+6q^4+8q^5+13q^6+17q^7+27q^8+ 35 q^9\ldots )\cr
q^{1\o 24}b_0^2&=
q^{1\o 3}(1+2q+3q^2+5q^3+8q^4+13q^5+19q^6+28q^7+41q^8+58q^9\ldots)\cr
 q^{1\o 24}b_2^2 & =
q^{1\o 12}(1+q+3q^2+4q^3+8q^4+11q^5+18q^6+25q^7+38q^8+52q^9\ldots).\cr}}

\subsec{Characters from Bethe's equations.}

In order to obtain the characters~\ctspf\ and \ctspa\ from the
formalism of Bethe's equation~\bethe, the order one computations
of~\rADMtwo~and~\raltwo~must be extended to order ${1\o M}$. We consider
the ferromagnetic and the anti-ferromagnetic cases separately, simpler
case first.
\vskip10pt
\no
{$\bullet$~ Anti-ferromagnetic characters:}

To extend the analysis of~\rADMtwo~to order ${1\o M}$ it is natural to use
the order one energies~\antifenone\
in the region $P\sim
{1\over M}$ where the linear form~\antipzero~holds. However, we must
in addition (i)
add a possible $P$-independent contribution of
order ${1\o M}$ to the energy and (ii) specify the allowed values of
$P_j^{\alpha}$, as we now explain.

 Both of these
questions are investigated in detail
in~\rKedMc. A principal result of that paper is that the ${1\o M}$ spectrum
decouples into a spectrum of right- and left-movers, namely
\eqn\antien{E-E_{GS}~=\sum_{\alpha=2s,-2s,ns}~~\sum_{h=r,l}
{}~~\sum_{j_{\alpha(h)}=1}^{m_{\alpha(h)}}
 e_{\alpha}(P_{j_{\alpha(h)}}^{\alpha(h)})}
with all $e_\alpha(P)=3|P|$, and (in the $Q=0$ sector)
\eqn\twos{ P_{j_{\alpha(h)}}^{\alpha(h)} ~=~ \pm {2\pi\o M} \Bigl[
 {1\o 2}\bigl(m_{ns(h)} + {m_{2s(h)}+m_{-2s(h)}\o 2}+1\bigr)+
 k_{j_{\alpha(h)}}^{\alpha(h)} \Bigr]}
for $\alpha=2s,-2s$, and
\eqn\ns{ P_{j_{ns(h)}}^{ns(h)} ~= ~\pm {2\pi\o M} \Bigl[
 {1\o 2}\bigl(m_{ns(h)} + m_{2s(h)}+m_{-2s(h)}+1\bigr)+
 k_{j_{ns(h)}}^{ns(h)} \Bigr]~~,}
where the $+,-$ applies to $h=r,l$, respectively,
and the $k_{j_{\alpha(h)}}^{\alpha(h)}$ are distinct non-negative
integers for each $\alpha(h)$.

It is significant that the lower limits on the three momentum ranges
depend on the number of quasi-particles present in the state. It is
this exclusion of states in the infrared that causes the specific heat
of this system to be less than that of 3 free fermions, namely less
than ${3\o 2}$.

{}From this order ${1\o M}$ energy spectrum we may construct the branching
functions $b_0^0$, $b_4^0$ and $b_2^0$ for $N=4$ by using~\twos\ and~\ns\
in~\Zsum\ with
$m_{\alpha(l)}=0$ and $m_{\alpha(r)}$
satisfying the following restrictions (where the subscript $r$ is
dropped for convenience):
\eqn\bzerozero{\eqalign{b_0^0:~~~~~~&
 m_{2s}+m_{-2s}{\rm \quad is~ even,~and\quad}
 m_{ns}+m_{-2s}+{m_{2s}+m_{-2s}\over 2} {\rm\quad is~ even};\cr
b_4^0:~~~~~~&m_{2s}+m_{-2s} {\rm \quad is~ even,~and\quad}
m_{ns}+m_{-2s}+{m_{2s}+m_{-2s}
\over 2}{\rm \quad is ~odd};\cr
b_2^0:~~~~~~&m_{2s}+m_{-2s}{\rm \quad is ~odd,~and\quad}
 m_{2s}<m_{-2s}.}}

The branching functions are now evaluated from
\eqn\branch{q^{c/24}b_m^0~=\sum_{m_{2s(r)},m_{-2s(r)},m_{ns(r)} }
  e^{-(E-E_{GS})/k_B T}~~,}
using the relation
\eqn\sumq{\sum_{N=0}^{\infty}Q_m(N)q^N = {q^{m(m-1)/2}\over (q)_m}}
where
$Q_m(N)$ is the number of distinct additive partitions of $N$
into $m$ non-negative integers.
We find that
\eqn\qzerob{
 \sum_{m_{2s},m_{-2s},m_{ns}=0}^{\infty}
{q^{{1\over4}(3m_{2s}^2+3m_{-2s}^2+4m_{ns}^2+4m_{ns}m_{2s}+4m_{ns}m_{-2s}
+2m_{2s}m_{-2s})}\over (q)_{m_{2s}}(q)_{m_{-2s}}(q)_{m_{ns}}}=q^{1/24}b_m^0}
with the $m_{\alpha}$ restricted by~\bzerozero.
The lhs is obtained directly from
Bethe's equation~\bethe. However, if we set
$m_{1}=m_{2s}$,
$m_2=m_{ns}$ and $m_{3}=m_{-2s}$, we see that
it is exactly the rhs of
\lpsum\ obtained by Lepowsky and Primc~\rlp,
 and thus the equality in \qzerob\ follows.

The sector $Q=\pm1$ is more complicated and for details the reader is
referred to~\rKedMc. The analysis there shows that
 each of the branching
functions $b_0^2$ and $b_2^2$ is represented in terms of
two types of
spectra with non-trivial lower bounds. The final result is that these
branching functions are given as  the sum of two 3-dimensional sums as follows:
\eqn\btwo{\eqalign{
 \sum_{{ m_1,m_2 ,m_3=0 \atop m_1+m_3~{\rm even}}}^\infty
 {q^{{\bf m}C_{3}^{-1}{\bf m}^t +m_1+m_2+m_3-{1\o 4}}\over
 (q)_{m_1} (q)_{m_2} (q)_{m_3}}  +
 \sum_{{ m_1,m_2 ,m_3=0 \atop m_1+m_3~{\rm odd}}}^\infty
 {q^{{\bf m}C_{3}^{-1}{\bf m}^t +{1\o 2}(m_1+m_3)}\over
 (q)_{m_1} (q)_{m_2} (q)_{m_3}} &= q^{{1\o 24}-{1\o 3}} b_0^2 \cr
 \sum_{{ m_1,m_2 ,m_3=0 \atop m_1+m_3~{\rm odd}}}^\infty
 {q^{{\bf m}C_{3}^{-1}{\bf m}^t +m_1+m_2+m_3-{1\o 4}}\over
 (q)_{m_1} (q)_{m_2} (q)_{m_3}}  +
 \sum_{{ m_1,m_2 ,m_3=0 \atop m_1+m_3~{\rm even}}}^\infty
 {q^{{\bf m}C_{3}^{-1}{\bf m}^t +{1\o 2}(m_1+m_3)}\over
 (q)_{m_1} (q)_{m_2} (q)_{m_3}} &= q^{{1\o 24}-{1\o 12}} b_2^2~.\cr}}
Unlike the case of the $Q=0$, the lhs's here are not of the form
\lpsum\ of~\rlp . Nevertheless, we have verified to order $q^{200}$
that the identities \btwo\ hold.

\vskip11pt
\no
{$\bullet$~ Ferromagnetic characters:}

The extension of the ferromagnetic order one spectrum~\ferroe ~to the
order ${1\o M}$ region is complicated by the degeneracy factor~\deg. At
order one this degeneracy may be thought of as additional
excitations  which must be included in the sum~\ferroe\
but have zero energy and zero momentum. However, at order ${1\o M}$ such
excitations can have  dispersion relations linear in $P$ just as long
as the number of allowed momentum states is finite as $M \rightarrow
\infty$. It is also not instantly obvious that the speed of sound of
these finite-momentum-range excitations should be the same as the
speed of sound of the quasi-particle of~\ferropzero. These questions
have been investigated in~\rDKMM~where we find that the characters
can be computed from the following expressions for the energies
\eqn\ferroen{E-E_{GS}~=\sum_{a = +,-2s,ns}~~\sum_{j_a=1}^{m_{a}}
 e_{a}(P_{j_a}^{a})~~,}
where   (in the $Q=0$ sector)
\eqn\pplus{ P_{j_+}^+ ~=~ {2\pi\o M} \Bigl[
 -{1\o 2}(m_- + m_{-2s}-1)+ k_{j_+}^+ \Bigr] }
\eqn\pmtwos{ P_{j_{-2s}}^{-2s}~ =~ {2\pi\o M} \Bigl[
 -{1\o 2}(m_- + m_{-2s}-1)+ k_{j_{-2s}}^{-2s} \Bigr]~~}
\eqn\pns{ P_{j_{ns}}^+ ~=~ {2\pi\o M} \Bigl[
 -{1\o 2}(m_{ns}+2m_- +2m_{-2s}-1)+ k_{j_{ns}}^{ns} \Bigr]~,}
with the $k_{j_{a}}^{a}$ distinct non-negative integers for each $a$,
which for $a=-2s,ns$
also have an upper bound,
\eqn\pmaxa{ k_{j_{-2s}}^{-2s}\leq m_-+m_{-2s}-1~~, ~~~~~~
 k_{j_{ns}}^{ns} \leq m_{ns}+ 2m_- +2m_{-2s}-1~~.}
Here
\eqn\mplus{m_{+}=2m_{ns}+3m_{-}+4m_{-2s}}
and
\eqn\ep{e_{a}(P)=3P~~,~~~~~~~a=+,-2s,ns.}
We emphasize that \ep\ differs from~\elin~in that $P$ occurs instead of
$|P|$, which is significant  since the lower limit in \pplus\
is in general negative. Note also that the number of states
allowed by \pmtwos-\pmaxa\ is finite as $M\to\infty$ for any given
$m_+$.

\medskip

Expressions for the characters $\chi_0$ and $\chi_3$ are constructed
using these rules with the further restriction
\eqn\evenodd{m_-\quad {\rm is}\quad {\rm even~(odd)~~~~~for}~~~
  \chi_0~~(\chi_3).}
The characters $\chi_{2/5}$ and $\chi_{7/5}$ may also be constructed
from these rules provided we add an additional term
\eqn\shift{{\pi v\over M}(m_{-}+m_{-2s}-1)}
to the energy, set $m_+=2m_{ns}+3m_-+4m_{-2s}-1$ and use the restrictions
\eqn\evenodd{m_-\quad {\rm is}\quad {\rm even~(odd)~~~~~for}~~~
  \chi_{2/5}~~(\chi_{7/5}).}

The evaluation of the sum for the characters is done
as in the anti-ferromagnetic case,
except that here,
because of the finite momentum ranges   dictated by \pmaxa,
it is not sufficient to use~\sumq. In addition, we
must introduce $Q_m(N;N')$, the number of partitions of $N\geq0$ into
$m$ non-negative integers which are smaller or equal to $N'>0$. Then
the sums may be simplified using~\rstan
\eqn\Qgen{ \sum_{N=0}^\infty Q_m(N;N') ~q^N ~=~ q^{{1\o 2} m(m-1)}~
  {N'+1 \atopwithdelims[] m}_q~~,}
where the $q$-binomial is defined (for integers $m,n$) by
\eqn\qbin{ {n \atopwithdelims[] m}_q ~=~\cases{ ~~{(q)_n \o (q)_m (q)_{n-m}}
  ~~~~~~~~& if ~~$0\leq m \leq n$ \cr  ~~0 & otherwise.\cr} }
Thus we may directly find, for example,
\eqn\chizero{q^{c/24} \chi_{0,3} ~= \sum_{m_{ns},m_{-2s},m_{0}=0\atop{\rm
restrictions}}^\infty {q^{F({\bf m})}\over(q)_{m_+}} {m_{-2s}+m_-
\atopwithdelims[] m_-}_q{2(m_-+m_{-2s})+m_{ns}\atopwithdelims[]m_{ns}}_q}
where
\eqn\qqff{F({\bf m}) = 2 m_{ns}^2 + 3 m_-^2 + 6 m_{-2s}^2 + 4 m_{ns}m_- + 6
m_{ns} m_{-2s} + 8 m_- m_{-2s}~,}
$m_+$ is given by~\mplus, and the restrictions on the sum are given
by~\evenodd.

For $Q=\pm1$ considerations similar to those of the anti-ferromagnetic case
give a representation of $\chi_{1/15}$ as the sum of five 3-fold sums
with an additional linear term in the exponent.

However, in all cases $Q=0,\pm1$ the structure of the result is much
more transparent if we set
\eqn\em{ m_1 = 2m_{ns}+3m_-+4m_{-2s}~,~~~~~m_2=2m_-+2m_{-2s}~,
 ~~~~~m_3 = m_-.}
Then we find the following set of results
\eqn\ferrores{\eqalign{
  \hat{\chi}_\Delta(q)~ =  \sum_{m_1,m_2,m_3\geq 0\atop{\rm
\scriptstyle  restrictions}}& q^{{1\o4} (2 m_1^2 +
2 m_2^2 + 2 m_3^2- 2 m_1 m_2 - 2 m_2 m_3)-{1\o2}L({\bf m})}\cr
& \times{1\o(q)_{m_1}} {{1\o2} ( m_1+m_3+u_2) \atopwithdelims[] m_2 }_q
{{1\o2}(m_2+u_3)\atopwithdelims[] m_3}_q
}}
where the quadratic form in the exponential is recognized as
${1\o 4}{\bf m}C_3 {\bf m}^t$ where $C_3$ is the Cartan matrix of
the Lie algebra $A_3$. The
restrictions differ according to the character, (below
{}~e=even, o=odd, and possibly several possibilities
for obtaining a given character are listed):

\medskip
\centerline{
\vbox{\offinterlineskip
\hrule
\halign{&
 \strut\quad\hfil#\quad\hfil\hfil\hfil\hfil\cr
$\Delta$\hfil&$m_1$\hfil&$m_2$\hfil&$m_3$\hfil&$u_2$\hfil&$u_3$
&$L({\bf m})$\cr
\noalign{\hrule}
$0$&e&e&e&0&0&0\cr
\noalign{\hrule}
$2/5$&o&e&e&1&0&1\cr
       &o&o&o&0&1&1\cr
\noalign{\hrule}
$7/5$&e&e&o&1&0&1\cr
       &e&o&e&0&1&1\cr
\noalign{\hrule}
$3$&o&e&o&0&0&6\cr
\noalign{\hrule}
$1/15$&o&e&o&2&0&$m_2+2$\cr
        &e&e&e&2&0&$m_2$\cr
        &e&o&o&1&1&$m_2$\cr
        &o&o&e&1&1&$m_2$\cr
        &$\{$e&o&e& & & \cr
        &+o&o&e$\}$&1&$-1$&$m_1-m_3$\cr
\noalign{\hrule}
$2/3$&e&e&o&1&0&$m_2+1$\cr
        &o&e&e&1&0&$m_2+1$\cr
        &$\{$e&o&e& & & \cr
        &+o&o&o$\}$&0&$-1$&$m_1-m_3+1$\cr}
\hrule}}

\medskip
There are a few comments to be made about this summary of
the results of~\rKedMc\ and \rADMtwo. Firstly, for $Q=\pm 1$ the
form~\lpsum\ for the anti-ferromagnetic characters and~\ferrores\ for the
ferromagnetic characters have not been derived from Bethe's
equation~\bethe\ but have been verified to hold to order $q^{200}$.
Secondly, the crucial factorization property~\zfac\ has only been shown
for the anti-ferromagnetic case. Thus the momentum restrictions and the
energy formula which give the characters by restricting to right-movers
only as in \branch\ do not seem to
be sufficient to give the full partition function
in the ferromagnetic case. This is still under
investigation, and the resolution presumably lies in the fact that the
limited range excitations~\pmtwos~and~\pns~can have many different
forms at the order of ${1\o M}$ which will all be degenerate at order 1.

\bigskip
\newsec{Generalizations}

The Lepowsky-Primc form \lpsum\ for the anti-ferromagnetic 3-state Potts
characters and the expressions for the ferromagnetic characters~\ferrores\
are written in terms of the Cartan and inverse Cartan matrix of $A_3$
and are extremely suggestive for generalizations.
We have recently conjectured such
generalizations~\rKKMMone \rKKMMtwo~for many conformal field theories,
including all those mentioned in sect.~3, and found that the conjectures
agree with the previously known results to order $q^{200}$ in many cases.
Furthermore, by reversing the process of the previous section, each of
the characters can be given an
interpretation in terms of
fermionic quasi-particles with momentum restrictions. We will here summarize
both these conjectures and other recent results for fermionic sum
representations.

\subsec{${(G_r^{(1)})_1\times(G_r^{(1)})_1\over (G_r^{(1)})_2}$~ where $G_r$ is
a simply-laced Lie algebra of rank $r$.}

Let us first define the general sum
\eqn\SofX{  S_B^Q(q) ~\equiv~ \sum_{m_1,\ldots ,m_n=0
\atop{\rm restrictions}}^{\infty} ~
    {q^{{1\over 2}{\bf m} B {\bf m}^t} \o (q)_{m_1} \ldots (q)_{m_n} }
    ~~,}
where $B$ is a real positive-definite $n\times n$ symmetric matrix, and
the restrictions are generically of the form
\eqn\crest{\sum_{\alpha=1}^n m_\alpha Q_\alpha \equiv Q~({\rm mod}~\ell) ~.}
The sum \SofX\ is the partition function of a set of $n$
types of (right-moving, say)
fermionic quasi-particles with momenta specified by
\eqn\Momdef{P^\alpha_{j_\alpha}~=~P_{{\rm min}}^\alpha({\bf m})
 +{2\pi\o M} k^\alpha_{j_\alpha}~,}
where the $k_{j_\alpha}^\alpha$ are distinct non-negative integers
for each $\alpha$ and
\eqn\Pmin{  P_{{\rm min}}^\alpha({\bf m})~ =~ {2\pi\o M}
 \Biggr[{1\o 2}
{}~+~ {1 \over 2}\sum_{\beta=1}^n \left( B_{\alpha\beta}
   -  \delta_{\alpha\beta} \right) m_\beta \Biggl]~. }
The interpretation of a restriction \crest\ is that each
quasi-particle of type $\alpha$ carries a $\ZZ_\ell$ charges $Q_\alpha$,
and so $S_B^Q$ is the partition function of the
sector of total charge $Q$.

To obtain  characters for the coset conformal field theory
{}~${(G_r^{(1)})_1\times(G_r^{(1)})_1\over (G_r^{(1)})_2}$~ we take $n=r$
and $B=2C_{G_r}^{-1}$, namely twice the inverse Cartan matrix of $G_r$.
The results in the various cases are as follows:

\medskip \no
${\bf G_r = A_n}$: ~This is the original case of Lepowsky and
Primc~\rlp :
the sum \SofX\ with $B=2C_{A_{N-1}}^{-1}$ is  \lpsum\ with $l=0$.
All the characters of the corresponding $\ZZ_{n+1}$-parafermionic
conformal field theory are given by~\lpsum.
We merely note here
that the linear shift term ${\bf A}_l\cdot{\bf m}$ of \lin\ can be obtained
from the form \lpsum\ with ${\bf A}=0$ by replacing in the quadratic form
$m_l$ by $m_l+{1\o 2}$.

\medskip \no
${\bf G_r= D_n~(n\geq 3)}$: The corresponding conformal field theories are
special points on the $c$=1 gaussian line (specified by the radius
$\sqrt{{n\o 2}}$ in the conventions of~\rGinsp ),
where the characters are given by
\eqn\fdef{
   f_{n,j}(q) ~=~ {q^{-1/24}\o (q)_\infty}
 ~\sum_{k=-\infty}^{\infty} q^{n(k+{j\o 2n})^2}~~, ~~~~~~~j=0,\ldots,n.}
The inverse Cartan matrix is
\eqn\qfDn{ \eqalign{ {\bf m} C_{D_n}^{-1} {\bf m}^t ~=~
  & \sum_{\alpha=1}^{n-2} \alpha m_\alpha^2 ~+~ {n\o 4}(m_{n-1}^2 ~+~ m_n^2)
  ~+~ 2\sum_{1\leq \alpha < \beta \leq n-2} \alpha m_\alpha m_\beta \cr
  &+ ~ \sum_{\alpha=1}^{n-2} \alpha m_\alpha(m_{n-1}+m_n)
  ~+~ {n-2\o 2} m_{n-1} m_n~~,\cr} }
and we obtain
\eqn\SQofDn{ S^Q_{D_n}(q) ~=~ q^{1/24} ~f_{n,nQ}(q) }
with $Q=0,1$,
when summation in \SofX\ is restricted to
\eqn\ResDn{m_{n-1}+m_n \equiv Q~({\rm mod}~2).}
Note that due to the coincidence $D_3=A_3$ the expressions \lpsum\ and
\SQofDn\ are related when $n=3$ by (cf.~\rKaWa\rKedMc )
{}~$S^0_{D_3}=S^0_{A_3}+S^2_{A_3}$~ and ~$S^1_{D_3}=2S^1_{A_3}$.

\medskip \no
${\bf G_r= E_6}$: ~Here the conformal field theory is the minimal model
${\cal M}(6,7)$ of central charge $c={6\o 7}$
with the $D$-series partition function.
With a suitable labeling of roots we have
\eqn\carEviinv{ C_{E_6}^{-1} ~=~ \pmatrix{4/3 & 2/3& 1& 4/3& 5/3& 2\cr
                                         2/3 & 4/3& 1& 5/3& 4/3& 2\cr
                                         1&  1& 2& 2& 2& 3\cr
                                         4/3& 5/3& 2& 10/3& 8/3& 4\cr
                                         5/3& 4/3& 2& 8/3& 10/3& 4\cr
                                         2& 2& 3& 4& 4& 6\cr} ~,}
and we find (cf.~\roc)
\eqn\SQofEvi{ S^0_{E_6}(q) ~=~ q^{c/24}
  ~[\chi^{(6,7)}_{1,1}(q)+\chi^{(6,7)}_{5,1}(q)] ~~,
   ~~~~S^{\pm 1}_{E_6}(q) ~=~ q^{c/24}~ \chi^{(6,7)}_{3,1}(q) ~~, }
with the restrictions
\eqn\ResEvi{m_1-m_2+m_4-m_5 \equiv Q~({\rm mod}~3).}

\medskip \no
${\bf G_r= E_7}$: ~The conformal field theory is ${\cal M}(4,5)$ of
central charge $c={7\o 10}$. Now
\eqn\carEviiinv{ C_{E_7}^{-1} ~=~ \pmatrix{3/2& 1& 3/2& 2& 2& 5/2& 3 \cr
                                         1& 2& 2& 2& 3& 3& 4 \cr
                                         3/2& 2& 7/2& 3& 4& 9/2& 6 \cr
                                         2& 2& 3& 4& 4& 5& 6 \cr
                                         2& 3& 4& 4& 6& 6& 8 \cr
                                         5/2& 3& 9/2& 5& 6& 15/2& 9 \cr
                                         3& 4& 6& 6& 8& 9& 12 \cr} }
and we find
\eqn\SQofEvii{ S^0_{E_7}(q) ~=~ q^{c/24} ~\chi^{(4,5)}_{1,1}(q) ~~,
   ~~~~S^{1}_{E_7}(q) ~=~ q^{c/24} ~\chi^{(4,5)}_{3,1}(q) ~~, }
when the restrictions  are
\eqn\ResEvii{m_1+m_3+m_6 \equiv Q~({\rm mod}~2).}

\medskip \no
${\bf G_r= E_8}$: ~The coset in this case is equivalent to the Ising
conformal field theory ${\cal M}(3,4)$ of central charge $c={1\o 2}$.
Here
\eqn\carEviiiinv{ C_{E_8}^{-1}~=~\pmatrix{2& 2& 3& 3& 4& 4& 5& 6 \cr
                                          2& 4& 4& 5& 6& 7& 8& 10 \cr
                                          3& 4& 6& 6& 8& 8& 10& 12 \cr
                                          3& 5& 6& 8& 9& 10& 12& 15 \cr
                                          4& 6& 8& 9& 12& 12& 15& 18 \cr
                                          4& 7& 8& 10& 12& 14& 16& 20 \cr
                                          5& 8& 10& 12& 15& 16& 20& 24 \cr
                                          6& 10& 12& 15& 18& 20& 24& 30 \cr} }
and, without any restrictions in the sum \SofX,
\eqn\SofEviii{ S_{E_8}(q) ~=~ q^{c/24} ~\chi^{(3,4)}_{1,1}(q) ~~.}
We further note that
if  $m_1$ in the quadratic form in \SofX\  is replaced
by $m_1-{1\o 2}$ then
one obtains
(up to a power of $q$)
$\hat{\chi}_{1,1}^{(3,4)}+\hat{\chi}_{1,2}^{(3,4)}$, and
similarly replacing $m_2$ by $m_2-{1\o 2}$ the combination
$\hat{\chi}_{1,1}^{(3,3)}+
\hat{\chi}_{1,2}^{(3,4)}+\hat{\chi}_{1,3}^{(3,4)}$ is obtained.

\subsec{The cosets of ${(G_r^{(1)})_{n+1}\over U(1)^r}$~.}

This case has been considered in~\rTer~and~\rKNS~where the identity
characters in the corresponding generalized parafermion conformal
field theory~\rgeptwo~are given by~\SofX\ (with suitable restrictions
on the summation variables) by taking
$B=C_{G_r} \otimes C_{A_n}^{-1}$, which is
explicitly written in a double index notation as
\eqn\Bab{B_{ab}^{\alpha \beta}=(C_{G_r})_{\alpha \beta}(C_{A_n}^{-1})_{ab}\quad
{}~~~~\alpha ,\beta =1,\ldots ,r,\quad a,b,=1,\ldots ,n.}
When $G_r = A_1$, this reduces to the result \lpsum\ of \rlp.

\subsec{The non-unitary minimal models ${\cal M}(2,2n+3)$.}

This case has been discussed in~\rFNO~and~\rNRT.
Here one takes $B=2(C'_n)^{-1}$, where $C'_n$ is the Cartan matrix of the
tadpole graph with $n$ nodes, namely it differs from $C_{A_n}$ only in one
entry which is $(C'_n)_{nn}=1$. The sum $S_B(q)$,
with no restrictions, gives the (normalized) character
$\hat{\chi}_{1,n}^{(2,2n+3)}(q)$ corresponding to the lowest dimension
in the theory. All the other characters are obtained~\rFNO~by adding
suitable linear terms to the quadratic form in \SofX, leading to the
full set of sums appearing in the Gordon-Andrews identities~\rgor\rand.

\subsec{Unitary minimal models ${\cal M}(p,p+1)=
 {(A_1^{(1)})_{p-2}\times(A_1^{(1)})_1\over (A_1^{(1)})_{p-1}}$~.}

For this and subsequent cases we must extend the form~\SofX\ to
\eqn\Snauq{ S_B{{\bf Q}\atopwithdelims[]{\bf A}}({\bf u}|q) ~\equiv~
 \sum_{{\bf m}\atop {\rm restrictions}}
  q^{{1\o 2}{\bf m} B {\bf m}^t -{1\o 2}{\bf A}\cdot{\bf m}}
  ~\prod_{a=1}^n ~
  { ({\bf m}(1-B)+{{\bf u}\over 2})_a \atopwithdelims[] m_a}_q~~,}
where ${\bf A}$ and
${\bf u}$ are $n$-dimensional vectors of
integers and the argument ${\bf Q}$ indicates
certain restrictions on ${\bf m}$ (such that, in particular,
the upper entries of the $q$-binomials are integers).
We note that if ~$u_a=\infty$~ then
${({\bf m}(1-B)+{{\bf u}\over 2})_a\atopwithdelims[]
m_a}_q={1\o (q)_{m_a}}$. Thus if all $u_a=\infty$ the form~\SofX\ is
obtained, while if only $u_1=\infty$ a form similar to~\ferrores\ is
obtained.

Generalizing the discussion leading to \chizero,
the sum \Snauq\ can be shown~\rKKMMtwo~to be the partition function of
a set of $n$ quasi-particles having the same dispersion relation
$e_{a}(P^a_{j_a})=vP^a_{j_a}$ for all $a=1,\ldots,n$,
and the $P^a_{j_a}$
{}~($j_a=1,2,\ldots,m_a$
with the $m_a$ restricted according to ${\bf Q}$)
obey the exclusion principle \fermi\ but are otherwise
freely chosen from the sets
\eqn\Pa{P^a_{j_a} \in \Bigl\{ P^a_{\rm min}({\bf m}),
  ~P^a_{\rm min}({\bf m})+{2\pi \o M},~\ldots,
  ~P^a_{\rm max}({\bf m}) \Bigr\}~.}
The vectors ${\bf P}_{\rm min,max}=\{P^a_{\rm min,max}\}$ here are
\eqn\Pmin{{\bf P}_{\rm min}({\bf m})
  ~ =~-{2\pi \over M}~{1\o 2} \Bigl({\bf m}(1-B)+{\bf A}-\rhob \Bigr)}
where $\rhob$ denotes the $n$-dimensional vector $(1,1,\ldots,1)$,
\eqn\Pmax{ P^a_{\rm max}({\bf m}) ~=~ -P^a_{\rm min}({\bf m})+
                  {2\pi \o M}({{\bf u}\o 2} -{ \bf A})_a~~,}
and we note that if some $u_a=\infty$ the corresponding
$P^a_{\rm max}=\infty.$

\medskip
For the present case of ${\cal M}(p, p+1)$ the
${\bf Q}$-restriction is taken to be $m_a\equiv Q_a$~(mod~2), and
\eqn\bmat{B~=~{1\o 2}C_{A_{p-2}}~,~~~~~\quad u_1=\infty ~.}
Defining
\eqn\QAurs{ \eqalign{ {\bf Q}_{r,s} ~ =~ (s-1)\rhob &+({\bf e}_{r-1}
  +{\bf e}_{r-3}+\ldots)
  + ({\bf e}_{p+1-s} +{\bf e}_{p+3-s}+\ldots)~~\cr}}
where $({\bf e}_a)_b=\delta_{ab}$ for $a=1,\ldots,p-2$ and 0 otherwise,
the conjecture for the (normalized) Virasoro characters \roc\
is~\rKKMMtwo
\eqn\main{ \eqalign{ \hat{\chi}^{(p,p+1)}_{~r,s}(q)~ &=~
  q^{-{1\o 4} (s-r)(s-r-1)}
  S_B{{\bf Q}_{r,s}\atopwithdelims[]{\bf e}_{p-s} }
  ({\bf e}_r +{\bf e}_{p-s}|q)~. \cr}}
Due to the symmetry $(r,s)$$\leftrightarrow$$(p-r,p+1-s)$ of the conformal
grid, another representation must also exist, namely
\eqn\mainalt { \hat{\chi}^{(p,p+1)}_{~r,s}(q)
 =~q^{-{1\o 4} (s-r)(s-r-1)}~
  S_B{{\bf Q}_{p-r,p+1-s}\atopwithdelims[]{\bf e}_{s-1} }
  ({\bf e}_{p-r}+{\bf e}_{s-1} |q)~~. }

\subsec{Cosets
 ~${(G_r^{(1)})_k\times(G_r^{(1)})_l\over (G_r^{(1)})_{k+l}}$~
 with $G_r$  simply-laced.}

In this case ~$B=C_{G_r}^{-1} \otimes C_{A_{k+l-1}}$, and
 the infinite entries of the vector ${\bf u}$
are $u^{\alp}_l$ for all
$\alp=1,\ldots,r$, in the double index notation used
in subsect.~5.2.

As an example with both $k$ and $l$ greater than 1, consider the case
$G=A_1$ with $l$=2, the resulting
series of theories labeled by $k$ being the unitary $N$=1 superconformal
 series whose characters are given
in~\rGKO . We find that the character
corresponding to the identity superfield
in these models is obtained by summing over
$m_1\in \ZZ$, $m_a \in 2\ZZ$ for $a=2,\ldots,k+1$.

Another example is the coset
{}~${(E_8^{(1)})_2 \times (E_8^{(1)})_1 \o (E_8^{(1)})_{3}}$~ of
central charge    ~$c= {21\o 22}$, which is
identified as the minimal model ${\cal M}(11,12)$
(with the partition function of the $E_6$-type).
The corresponding sum \Snauq, with ${\bf A}$=0, $u^{\alp}_2$=0
for all $\alp=1,\ldots,8$, and all 16 summations running over all
non-negative integers, gives
{}~$\hat{\chi}_{1,1}^{(11,12)}(q)+q^8 \hat{\chi}_{1,7}^{(11,12)}(q)$, which
is the (extended) identity character of this model.

\subsec{Non-unitary minimal models ${\cal M}(p,p+2)$~ $(p$ odd $)$.}

The normalized character
$\hat{\chi}^{(p,p+2)}_{(p-1)/2,(p+1)/2}(q)$~  (see \roc) corresponding to
the lowest conformal dimension
{}~$\Delta^{(p,p+2)}_{(p-1)/2,(p+1)/2}=-{3\o 4p(p+2)}$~ in this model
is given by \Snauq\ with ~$B={1\o 2}C_{(p-1)/2}'$ (where
$C_n'$ is defined in subsect.~5.3), ~${\bf A}$=0,
$u_1$=$\infty$ and $u_a$=0 for $a=2,\ldots,{p-1\o 2}$,
and the $m_a$ are summed over all even non-negative integers.

\subsec{Minimal models ${\cal M}(p,kp+1)$.}

For $k$=1 these models are the ones considered in sect.~5.4,
while for $p$=2 they were discussed in sect.~5.3. Here we
consider the general case.
The character
$\hat{\chi}^{(p,kp+1)}_{1,k}(q)$ corresponding to the lowest conformal
dimension in the model
is obtained from \Snauq\ with $B$ a $(k+p-3)\times(k+p-3)$
matrix whose nonzero
elements are given by ~$B_{ab}=2(C_{k-1}'^{-1})_{ab}$ ~and
{}~$B_{ka}$=$B_{ak}$=$a$~ for ~$a,b=1,2,\ldots,k-1$, and
{}~$B_{ab}={1\o 2}[(C_{A_{p-2}})_{ab}+(k-1)\delta_{ak}\delta_{bk}]$~ for
{}~$a,b=k,k+1,\ldots,k+p-3$.
Summation is restricted to even non-negative integers
for $m_k,\ldots,m_{k+p-3}$,
the other $m_1,\ldots,m_{k-1}$ running over all non-negative
integers, and
{}~$u_a$=$\infty$ for $a=1,\ldots,k$ and 0 otherwise.

The case $p$=3 is special in that
the fermionic sums are of the form \SofX\ for any $k$.
A slight modification of the matrix $B$ appropriate
for ${\cal M}(3,3k+1)$,
namely just setting $B_{kk}={k\o 2}$ while leaving all other elements
unchanged, gives the normalized
character $\hat{\chi}_{1,k}^{(3,3k+2)}$~ of ${\cal M}(3,3k+2)$.

\subsec{Unitary $N$=2 superconformal series.}

Expressions for the characters of these models,
of central charge ~$c={3k \o k+2}$~ where $k$ is a positive integer,
can be found in~\rgepth . The identity character,
given by ~$\chi_0^{0(0)}(q)+\chi_0^{0(2)}(q)$~ in the notation
of~\rgepth , can be obtained from \Snauq\ if one takes
{}~$B={1\o 2}C_{D_{k+2}}$, ~$u_k$=$\infty$ ~(in the basis used in \qfDn)
and all other $u_a$ set to zero, and $m_{k+1},m_{k+2}$ running
over all non-negative integers while all other $m_a$ summed only over
the even non-negative integers.

\subsec{$\ZZ_N$ parafermions.}

The characters of these models are the branching functions $b^l_m$
given by \blm, or by the fermionic representation \lpsum\ of~\rlp.
In sect.~4.4 we found another fermionic representation  for
the case $N$=3 which coincides with the minimal model ${\cal M}(5,6)$
with the $D$-series partition function.
(The $b^l_m$ in this case are linear combinations of the
$\chi_\Delta$ of \ferrores.)
Here we generalize the latter form to arbitrary $N$. For instance,
$b^0_0$ is obtained from \Snauq\ by setting
{}~$B={1\o 2}C_{D_{N}}$, ~$u_{N}$=$\infty$ ~(in the basis used in \qfDn)
and all other $u_a$ set to zero, and $m_{N-1},m_{N}$ running
over all non-negative integers such that $m_{N-1}+m_{N}$ is even,
while all other $m_a$ are restricted to be even.

\bigskip
\newsec{The $q \rightarrow 1$ behavior}

We  can now return to the discussion of specific heat and
conformal field theory of sections 2 and 3 by computing the effective
central charge~\Zasymp~directly from the $q$-series \Snauq.
The computation will use the steepest descent method of~\rRS~and~\rNRT.
We follow closely the presentation of~\rKKMMtwo.

It is easily seen that the $q \rightarrow 1$ behavior of \Snauq\ is
independent of the restrictions ${\bf Q}$ and the linear terms ${\bf A}$.
This is consistent with the $k$-independence of \chiasy.
Thus
without loss of generality we set ${\bf A}=0$ and let all the
sums run from
$0$ to $\infty$.
The resulting unrestricted sum will be denoted by $S_B({\bf u}|q)$.

Let $q=e^{2\pi i\tau}$ and
$\tilde{q}=e^{-2\pi i/\tau}$, with Im$\tau>0$. Then if
the coefficients in the series for $S_B({\bf u}|q)=\sum s_M q^M$
behave for large $M$ like $s_M \sim e^{2\pi\sqrt{\gamma M/6}}, ~\gamma>0$,
the series $S_B({\bf u}|q)$ diverges like
\eqn\SBlim{ S_B({\bf u}|q) ~\sim ~ \tilde{q}^{-\gamma/24}~~~~~~~~~
  {\rm as} ~~~q\to 1^-  ~. }
Here $\gamma$ must equal the effective central charge \ctilde\
of the corresponding conformal field theory.

The large $M$ behavior of $s_M$ is found by considering
\eqn\sm{ s_{M-1}~=~\oint {dq \o 2\pi i} ~q^{-M} ~S_B({\bf u}|q) ~=~
  \sum_{{\bf m}\geq {\bf 0}} \oint {dq \o 2\pi i}~
                 q^{-M} ~S_B^{\bf m}({\bf u}|q)~~,}
where the contour
of integration is a small circle around 0. The behavior of the integral
is now analyzed using a saddle point approximation. We first approximate
\eqn\lnSm{ \eqalign{ \ln  \Bigl(q^{-M} & S_B^{\bf m}({\bf u}| q)\Bigr)
  ~\simeq~ \bigl({1\o 2}{\bf m} B {\bf m}^t -M\bigr)\ln q \cr
  + & ~\sum_{a=1}^n
 \left( \int_0^{ ({\bf m}(1-B)+{{\bf u}\o 2})_a}-
        \int_0^{(-{\bf m}B+{{\bf u}\o 2})_a} -
        \int_0^{m_a} \right) dt ~\ln (1-q^t) \cr} }
for large ${\bf m}$, and set the derivatives of this expression
with respect to the $m_a$ to zero in order to find the saddle point.
Introducing ~$x_a={(1-w_a)v_a \o 1-v_a w_a}$~
and ~$y_a={1-w_a \o 1-v_a w_a}$~ where
{}~$v_a = q^{m_a}$~ and ~$w_a=q^{(-{\bf m}B+{{\bf u}\o 2})_a}$,
these extremum conditions reduce to
\eqn\xyeq{ 1-x_a ~=~ \prod_{b=1}^n x_b^{B_{ab}}~,~~~~~~~~
           1-y_a ~=~ \sigma_a \prod_{b=1}^n y_b^{B_{ab}}~~,}
where we define ~$\sigma_a$=0~ if ~$u_a$=$\infty$~ and 1 otherwise,
ensuring ~$y_a$=1 ~for ~$u_a$=$\infty$.

At the extremum point with respect to the $m_a$ we have
\eqn\lnSmext{ \eqalign{
 \ln \bigl( q^{-M} S_B^{\bf m}( &{\bf u}|q) \bigr) \Bigl|_{{\rm ext}} ~\simeq~
  -M\ln q \cr
 +~{1\o \ln q}\Biggl\{ & {1\o 2}\ln{\bf v} ~B~ \ln{\bf v}^t
 -\sum_{a=1}^n \bigl[{\cal L}(1-v_a)+{\cal L}(1-w_a)-{\cal L}(1-z_a)\bigr] \cr
 &-{1\o 2}\bigl[\ln{\bf v}\cdot\ln(1-{\bf v})+
               \ln{\bf w}\cdot\ln(1-{\bf w})-
               \ln{\bf z}\cdot\ln(1-{\bf z})\bigr] \Biggr\}~~ \cr} }
with ~$(\ln {\bf v})_a=\ln v_a$~ and
{}~$z_a = v_a w_a$, where
\eqn\dilog{ {\cal L}(z) = -{1\o 2} \int_0^z dt \left[
   {\ln t \o 1-t} + {\ln (1-t) \o t} \right]
  = -\int_0^z dt~{\ln (1-t)\o t}+{1\o 2}\ln z \ln(1-z)  }
is the Rogers dilogarithm function~\rLewin.
Now using \xyeq\ we see that the first term inside the braces in
\lnSmext\ cancels against the last.
Then using the five-term relation for the
dilogarithm~\rLewin\
\eqn\vterm{ {\cal L}(1-v)+{\cal L}(1-w)-{\cal L}(1-v w) ~=~
   {\cal L}(1-x)-{\cal L}(1-y)~~,}
where ~$x={(1-w)v \o 1-v w}$~ and ~$y={1-w \o 1-v w}$, we obtain
\eqn\lnSmexta{
 \ln\left(q^{-M} S_B^{\bf m}({\bf u}|q)\right)\Bigl|_{{\rm ext}} ~\simeq~
 -M \ln q - {\pi^2 \tilde{c} \o 6\ln q} ~~}
with
\eqn\ctildetw{ \tilde{c} ~=~ {6\o \pi^2} \sum_{a=1}^n
    \left[ {\cal L}(1-x_a)-{\cal L}(1-y_a) \right]~~.}
Finally  the value of $q$ at the saddle point is determined by extremizing
\lnSmexta\ with respect to $q$,
which leads to ~$s_M \sim e^{2\pi\sqrt{\tilde{c}M/6}}$~ and
consequently to \SBlim\ with $\gamma=\tilde{c}$ of~\ctildetw.

\medskip
This computation of the $q \rightarrow 1$ behavior of \Snauq\ is
completely general in that it is valid for all matrices $B$, and
presumably for an arbitrary $B$ no simplification
of~\xyeq\ and \ctildetw\ is possible. Nevertheless, for the conformal
field theories considered in sect.~5 there is one final
simplification which occurs. Namely, there is a remarkable set of
sum rules for
the dilogarithms~\rBR \rKun \rKR -\rFrenk~which
reduces~\ctildetw\ to rational
numbers. These sum rules must be regarded as a vital
piece of the theory, but are outside the scope of this
article and we refer the reader to the original papers for details.

Finally, it must be pointed out that for the models corresponding to
the conformal field theories of sect.~5 the specific heats
have been derived in a completely
independent fashion using the thermodynamic Bethe ansatz~\rBR \rKun \rKR
\rTBA~which
uses the definition of specific heat discussed in sect.~2. The
agreement of these two procedures establishes the one length-scale
scaling discussed in sect.~3.

\bigskip
\newsec{Discussion}

It is clear from sect.~5 that the existence of fermionic quasi-particle
representations for conformal field theory characters is a very general
feature which goes beyond
the specific models discussed
in sect.~4, where these representations were obtained from the
spectrum of the hamiltonian. In these representations the focus is
on the momentum selection rules \Pa-\Pmax. On the other hand, in
most of the previously known
expressions for the characters, obtained using conformal field
theory or representation theory methods, the focus is on
the modular transformation properties of the characters.
It would be interesting to directly relate these two aspects.

This question can be made explicit by focusing on the quadratic-form
matrix $B$ of~\Snauq. If this matrix is considered as coming from the
momentum restrictions for the fermionic quasi-particles there appears
to be nothing to distinguish one matrix $B$ from another. However,
from the point of view of conformal field theory the general
form~\Snauq\ can only represent a character if it is possible to find
some (possibly fractional) power of $q$ which, when multiplied by the
$q$-series \Snauq, gives a function which transforms properly under the
modular group. The mathematical structure of these $q$-series
cannot be said to be
fully
understood until these modular properties are
found directly from the series, which generalize the sum-side of the
Rogers-Ramanujan identities.

\medskip
A further property of great importance is the fact that there are
often several completely different fermionic $q$-series
representations for the same conformal field theory
characters. As a particular example, we
note that the
characters obtained as \ferrores\ from
the study of the ferromagnetic 3-state Potts hamiltonian
can also be written in the
Lepowsky-Primc form~\lpsum\
with $N$=3. More generally, the representations of the
$\ZZ_N$-parafermion
characters of section 5.9 and~\lpsum~are of different forms with
different quasi-particle interpretations, but nevertheless they are equal.
This is
representative of
a general phenomenon. A full discussion is beyond the scope of this
article, but we remark that
these inequivalent fermionic representations
for the characters are related to different
integrable perturbations of the model.

\medskip
Finally, there is the question of obtaining proofs of the several
conjectures of sect.~5.
One method
is to find certain finitizations of the $q$-series
in question into polynomials, whose properties can then be studied
using recursion relations.
Such a finitization exists
for the
characters of the unitary minimal
models \main, which matches the finitization of
the Rocha-Caridi formula \roc\ employed by Andrews,
Baxter and Forrester~\rABF~in their corner transfer matrix analysis
of the underlying RSOS models.
The details will be presented elsewhere.

\medskip

\bigskip
\bigskip \no
{\bf Acknowledgements}
\medskip
One of us (BMM) is deeply grateful to J.M.~Maillard for the opportunity
of participating in the conference ``Yang-Baxter Equations in Paris''
and for the great hospitality extended at this conference. He also wishes to
thank R.J.~Baxter for hospitality extended at the Isaac Newton
Institute where part of this work was performed. We also wish to acknowledge
useful discussions with
G.~Albertini, R.J.~Baxter, V.V.~Bazhanov, J.L.~Cardy, P.~Fendley,
C.~Itzykson,
A.~Kuniba, J.~Lepowsky, T.~Nakanishi, P.A.~Pearce,
L.A.~Takhtajan,
A.B.~Zamolodchikov and
Al.B.~Zamolodchikov.
The work of RK and BMM is
partially supported by the National Science Foundation under grant
DMR-9106648. The work of TRK is supported by NSERC and the Department
of Energy grant
DE-FG05-90ER40559, and that of EM under NSF grant 91-08054.

\vfill

\eject
\listrefs

\vfill\eject

\bye\end